\documentclass{ws-procs975x65}

\def\beq{\begin{equation}}
\def\eeq{\end{equation}}

\begin{document}

\title{Highly super-Chandrasekhar white dwarfs in an extensive GRMHD framework}
\author{Upasana Das}

\address{JILA, University of Colorado, \\ 
440 UCB, Boulder, CO 80309, USA\\
Email: upasana.das@jila.colorado.edu}

\author{Banibrata Mukhopadhyay}

\address{Department of Physics, Indian Institute of Science,\\
Bangalore 560012, India\\
E-mail: bm@physics.iisc.ernet.in}

\begin{abstract}
Our consistent effort to unravel the mystery of super-Chandrasekhar white dwarfs (WDs),
by exploiting the potential of magnetic fields, has brought this topic 
considerable attention. This is also evident from the recent surge in the 
corresponding literature. In the present work, 
by means of full-scale general relativistic
magnetohydrodynamic (GRMHD) numerical analysis, we confirm the existence of 
stable, highly magnetized, significantly super-Chandrasekhar WDs having mass 
exceeding $3$ solar mass. We have explored various possible field configurations, namely, poloidal, 
toroidal and mixed, by self-consistently incorporating the departure from spherical 
symmetry induced by a strong magnetic field. Such super-Chandrasekhar WDs can be ideal 
progenitors of peculiar, over-luminous type Ia supernovae. 
\end{abstract}


\bodymatter


\section{Introduction}

With the aim of obtaining a fundamental basis behind 
the formation of super-Chandrasekhar white dwarfs (WDs), 
Mukhopadhyay and his collaborators \cite{kundu,prd12,prl13,apjl13,jcap14,mg13} initiated the exploration of highly magnetized 
WDs and their new mass-limit, significantly exceeding the Chandrasekhar limit of $1.44M_\odot$.\cite{chandra35}
These WDs are ideally suited to be the progenitors of peculiar, overluminous, type Ia supernovae,
e.g. SN~2003fg, SN~2006gz, SN~2007if, SN~2009dc,\cite{nature,scalzo}
which are best explained by invoking the explosion of super-Chandrasekhar 
WDs having mass $2.1-2.8M_\odot$. Along with the fact that several 
WDs have been discovered with surface fields $10^5-10^9$ G, it has also been known that 
magnetized WDs tend to be more massive than their non-magnetized counterparts.\cite{kepler} 
Such observations motivate the theoretical investigation of
the effect of a strong interior magnetic field on the mass of a WD.

In this context, we mention that our previous attempts at obtaining highly magnetized super-Chandrasekhar WDs, assumed 
{\it a priori} spherical symmetry. While that may indeed be the case 
for certain magnetic field geometries, in general, highly magnetized WDs tend to be deformed 
due to magnetic tension. With each new step we have scientifically progressed towards a more rigorous model --- starting from a 
simplistic Newtonian, spherically symmetric, constant field model and culminating in a model with
self-consistent departure from spherical symmetry by general relativistic
magnetohydrodynamic (GRMHD) formulation, which we explain in the present work (also, see Ref. 11).
We appropriately modify the {\it XNS} code,\cite{xns1,pili} which has so far been used 
only to model strongly magnetized neutron stars, to compute equilibrium configurations of static, 
strongly magnetized WDs in the GR framework, for the first time in the literature 
to the best of our knowledge.

\section{Numerical set-up}
\label{setup}

For a detailed description of the underlying GRMHD equations, the magnetic field geometries, 
the numerical technique employed by the {\it XNS} code and the values of various code parameters, 
we refer the readers to Refs. 11-13.

We construct axisymmetric WDs in spherical polar coordinates $(r,\theta,\phi)$, 
to self-consistently account for the deviation from spherical symmetry due to a strong 
magnetic field, which generates an anisotropy in the 
magnetic pressure.\cite{monica} A uniform computational grid is used along 
both the radial $r$ and polar $\theta$ co-ordinates, the number of grid points 
being typically $N_r=500$ and $N_\theta=100$ respectively. Even 
higher resolution runs (for e.g. with $N_r=1000$ and $N_\theta=500$) require more computational 
time but do not lead to any significant change in the results.

In this work, we focus on the equilibrium solutions of high density, magnetized, 
relativistic WDs, which can be 
described by a polytropic equation of state (EoS) $P=K\rho^\Gamma$, where $P$ is the pressure 
and $\rho$ the density, such that the adiabatic index $\Gamma \approx 4/3$ 
and the constant $K$ is same as that obtained by Chandrasekhar.\cite{chandra35} 
Hence, we neglect the possible effect of Landau quantization on the above EoS which could arise due to a strong magnetic field 
$B>B_c$, where $B_c=4.414\times10^{13}$ G, is the critical magnetic field.\cite{prd12}
We recall that the maximum number of Landau levels $\nu_m$ occupied by electrons in the presence of a magnetic field 
is given by equation (10) of Ref. 1. The range of central density and maximum 
magnetic field strength inside the WDs considered in this work are 
$10^{10} \lesssim \rho_c \lesssim 10^{11}$ gm/$\rm cm^3$ and 
$10^{13} \lesssim B_{\rm max} \lesssim 10^{15}$ G respectively. Consequently, 
$\nu_m \gtrsim 20$ for this range of $\rho_c$ and $B_{\rm max}$, which is large enough not to
significantly modify the value of $\Gamma$ we choose, hence justifying our assumption.

\section{Results with different magnetic field configurations}

We now explore the effect of various magnetic field geometries on the structure and properties of WDs. 
For a fiducial model, we choose a non-magnetized WD with $\rho_c=2\times 10^{10}$ gm/$\rm cm^3$. 
It has a baryonic mass $M_0 = 1.416 M_\odot$ and 
equatorial radius $R_{eq}=1221.94$ km, and is perfectly spherical 
with $R_p/R_{eq}=1$, $R_p$ being the polar radius (note that for the definitions 
of all global physical quantities characterizing the solutions in this work, 
we refer to Appendix B of Ref. 13).

\def\figsubcap#1{\par\noindent\centering\footnotesize(#1)}
\begin{figure}[h]%
\begin{center}
  \parbox{2.1in}{\includegraphics[width=2in]{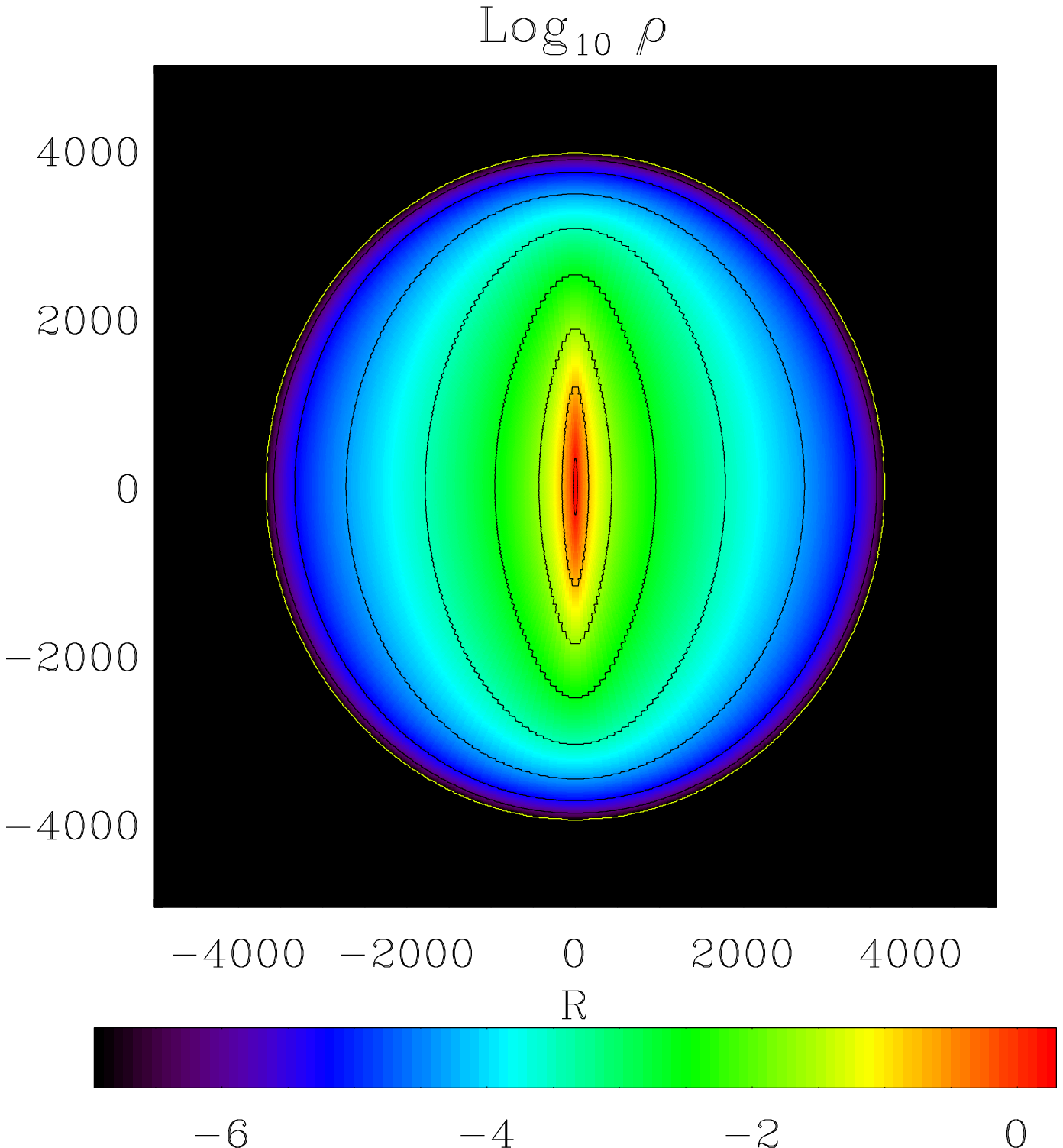}\figsubcap{a}}
  \hspace*{4pt}
  \parbox{2.1in}{\includegraphics[width=2in]{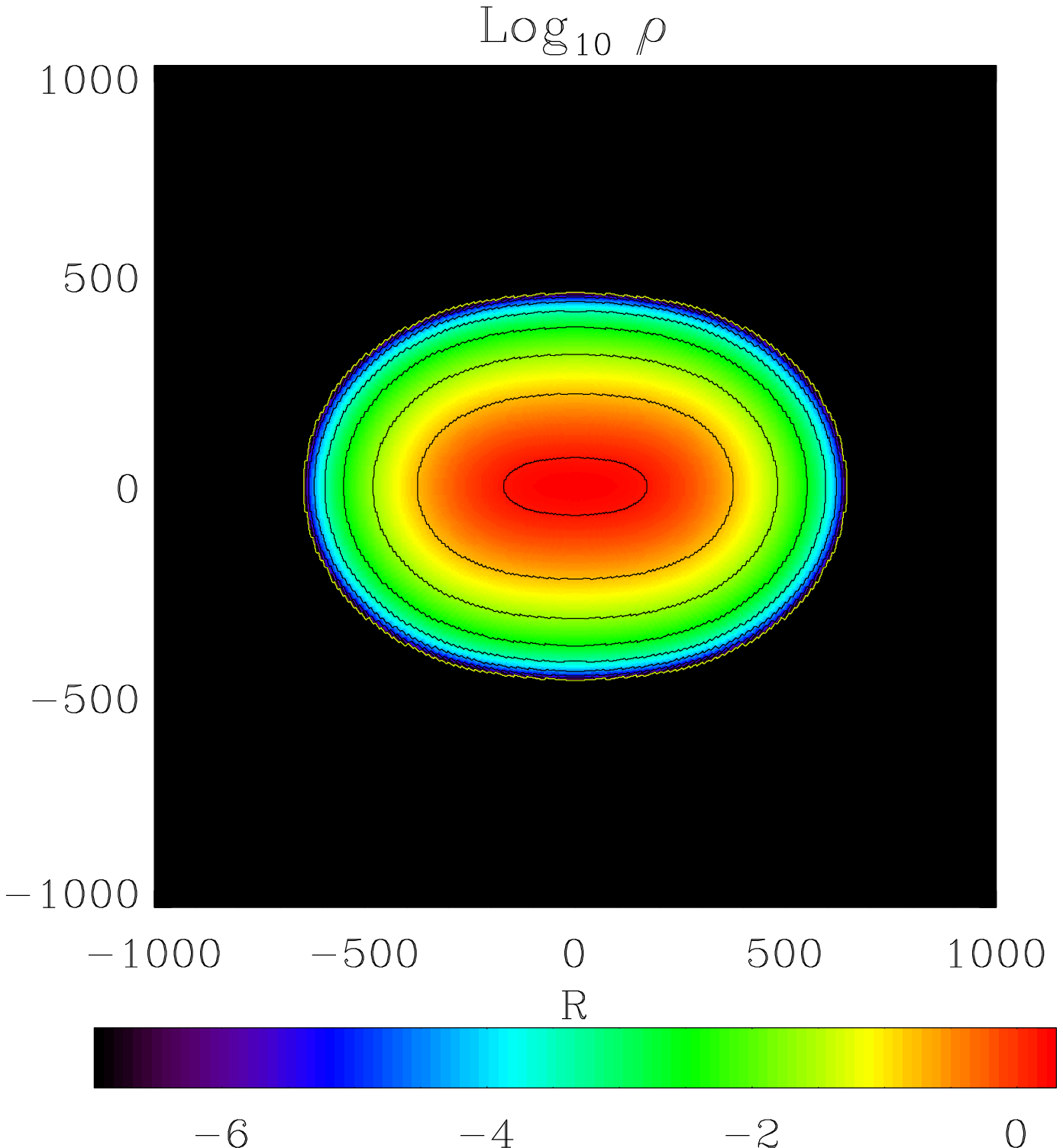}\figsubcap{b}}
\vspace*{4pt}
  \parbox{2.1in}{\includegraphics[width=2in]{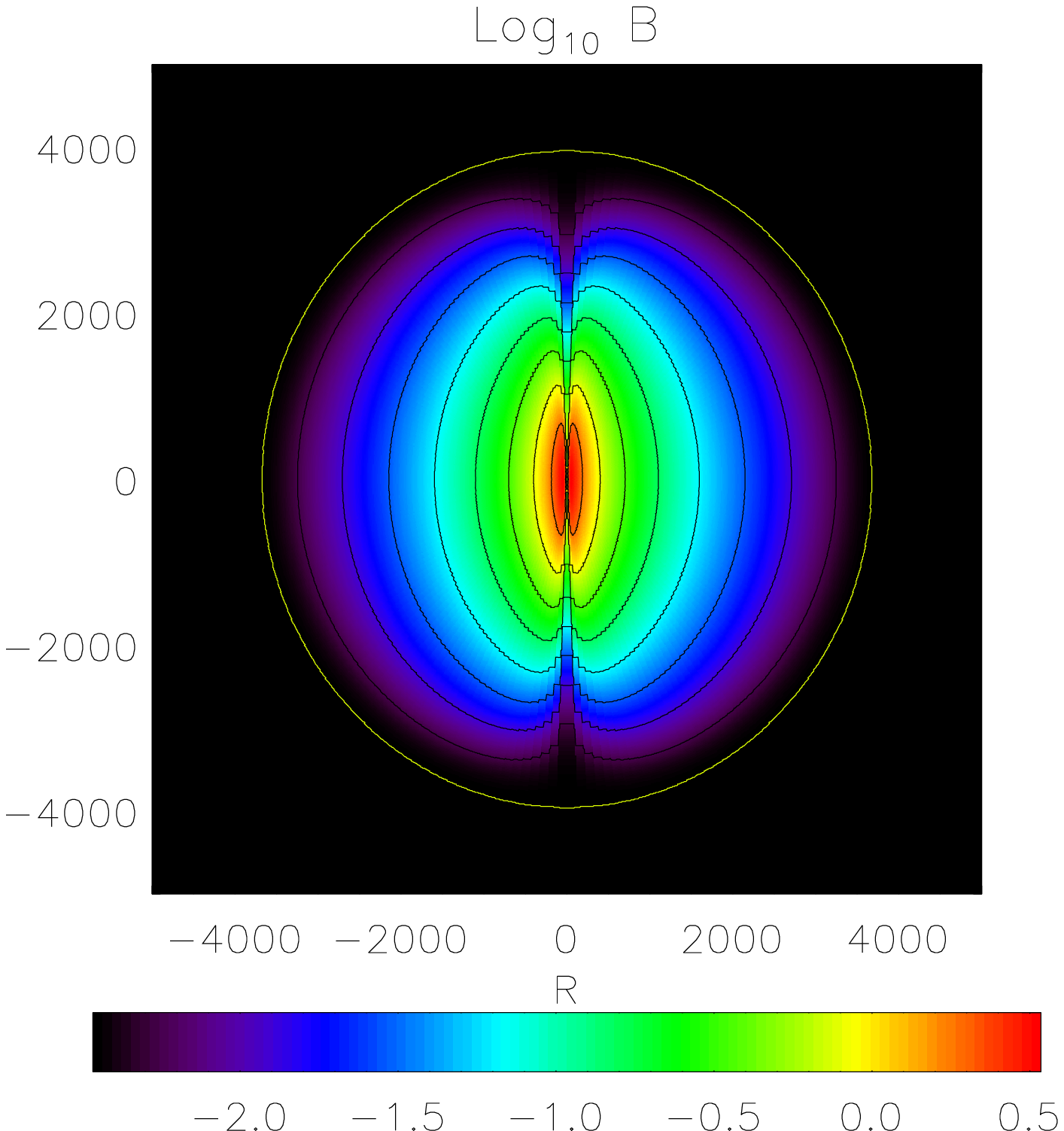}\figsubcap{c}}
\hspace*{4pt}
  \parbox{2.1in}{\includegraphics[width=2in]{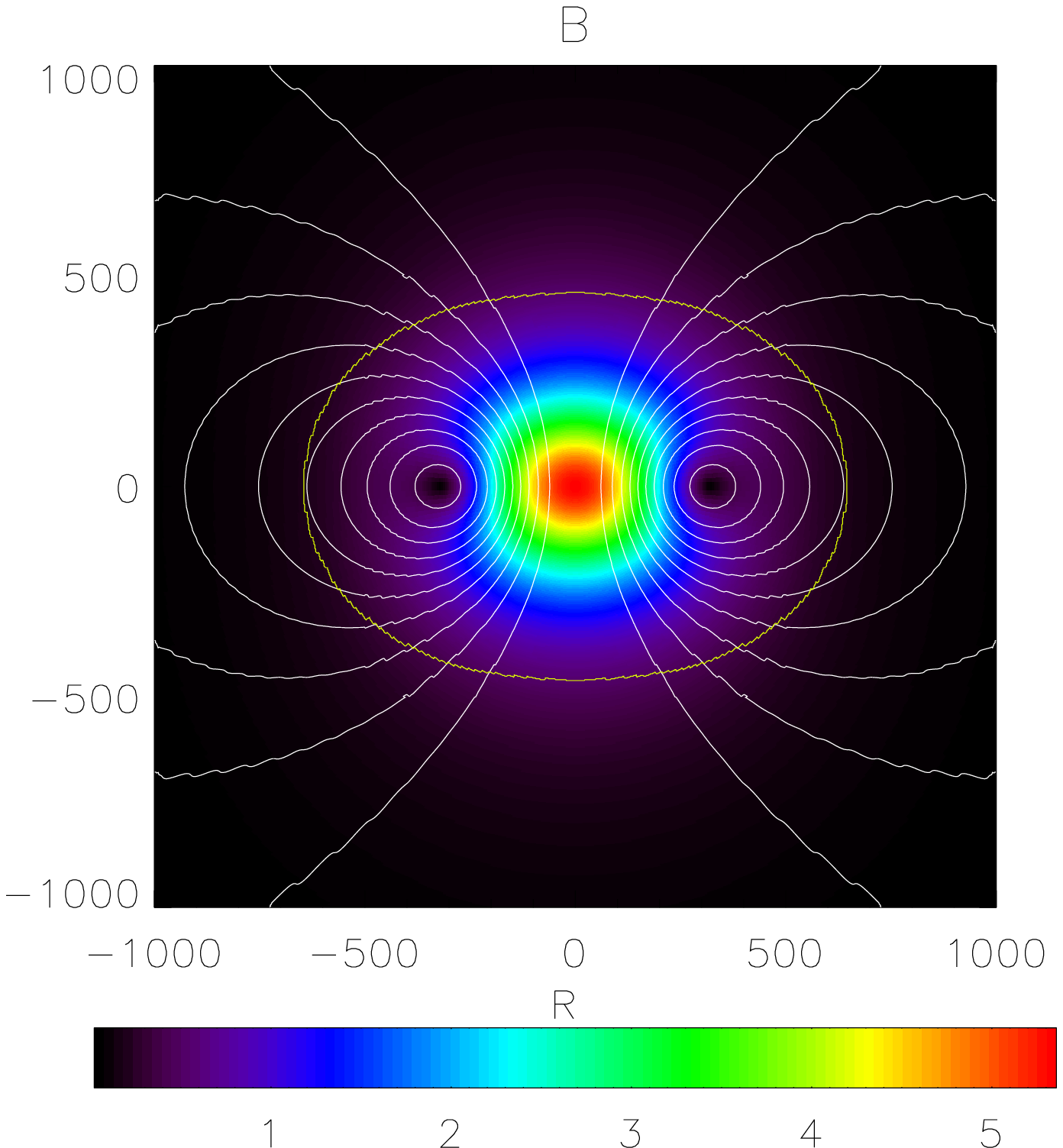}\figsubcap{d}}
  \caption{Purely toroidal configuration: iso-contours of (a) baryonic density and (c) magnetic field strength. 
Purely poloidal configuration: iso-contours of (b) baryonic density and (d) magnetic field strength, 
superimposed with dipolar magnetic field lines (in white). R is in units of $GM_\odot/c^2=1.476$ km, $\rho$ in units of $10^{10}$ gm/cm$^3$
and $B$ in units of $10^{14}$ G. The yellow curve in each panel represents the stellar surface.}%
  \label{poltor}
\end{center}
\end{figure}

\begin{figure}[h]
\begin{center}
\includegraphics[width=5.5in]{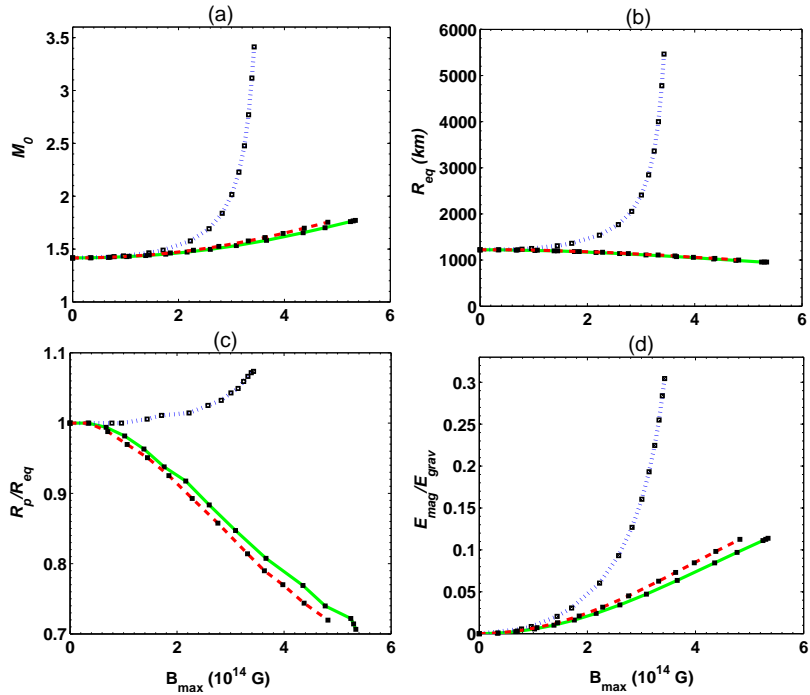}
\end{center}
\caption{Equilibrium sequences of magnetized WDs with fixed $\rho_c=2\times 10^{10}$ gm/$\rm cm^3$. (a) $M_0$, (b) $R_{eq}$, (c) $R_p/R_{eq}$ 
and (d) $E_{mag}/E_{grav}$, as functions of $B_{\rm max}$. The solid (green), dotted (blue) and dashed (red) 
curves represent, respectively, 
WDs having purely poloidal, purely toroidal, and twisted torus field configurations. $M_0$ is in units of $M_\odot$. 
The filled boxes represent individual WDs.}
\label{rhocf}
\end{figure}

Figures \ref{poltor}(a) and (c) portray the distribution of baryonic density and 
magnetic field strength respectively, for the fiducial WD 
having a purely toroidal magnetic field configuration $\vec{B}= B_\phi \hat{\phi}$. 
The maximum magnetic field strength inside this WD is 
$B_{\rm max}=\sqrt{B_\phi B^\phi} =3.41 \times 10^{14}$ G. Interestingly, this is a highly 
super-Chandrasekhar WD, having $M_0=3.413M_\odot$. Note that the value of the surface magnetic field does not
affect this result, as long as it is $\lesssim 10^{11}$ G, which is satisfied in this case. 
In this context, we 
mention that the detection of very high surface magnetic fields $\gtrsim 10^9$ G is 
very difficult due to the featureless spectrum.\cite{kepler}
Very importantly, the ratio of the total magnetic energy to the total gravitational 
binding energy, $E_{mag}/E_{grav}=0.3045$ (which is much $<1$). 
WDs with even smaller $E_{mag}/E_{grav}$ are also found to be highly super-Chandrasekhar 
(see Figs. \ref{rhocf}a and d). This argues for the WDs to be stable (see, e.g. Ref. 15). 
The radii ratio, $R_p/R_{eq}=1.074$ (which is slightly $> 1$), indicates a net prolate deformation 
in the shape caused due to a toroidal field geometry. 
Figure \ref{poltor}(a) shows that although the central iso-density contours are compressed 
into a highly prolate structure, the outer layers expand, giving rise to an 
overall quasi-spherical shape. Interestingly, this justifies the earlier 
spherically symmetric assumption in computing models of at least certain strongly 
magnetized WDs.\cite{prd12,prl13,jcap14}


Figures \ref{poltor}(b) and (d) show the distribution of baryonic density and 
magnetic field strength superimposed by magnetic field lines respectively, 
for the fiducial WD having a purely poloidal magnetic field configuration $\vec{B}=B_r \hat{r} + B_\theta \hat{\theta}$. 
The maximum magnetic field strength attained at its center is $B_{\rm max}=\sqrt{B_r B^r + B_\theta B^\theta} = 5.34\times 10^{14}$ G, 
which also leads to a significantly super-Chandrasekhar WD having $M_0=1.771M_\odot$. The WD is 
highly deformed with an overall oblate shape and 
$R_p/R_{eq}=0.7065$, which is expected to be stable because its $E_{mag}/E_{grav}=0.1138$, 
which is very much $<1$.\cite{ost} 

\def\figsubcap#1{\par\noindent\centering\footnotesize(#1)}
\begin{figure}[h]%
\begin{center}
  \parbox{2.1in}{\includegraphics[width=2in]{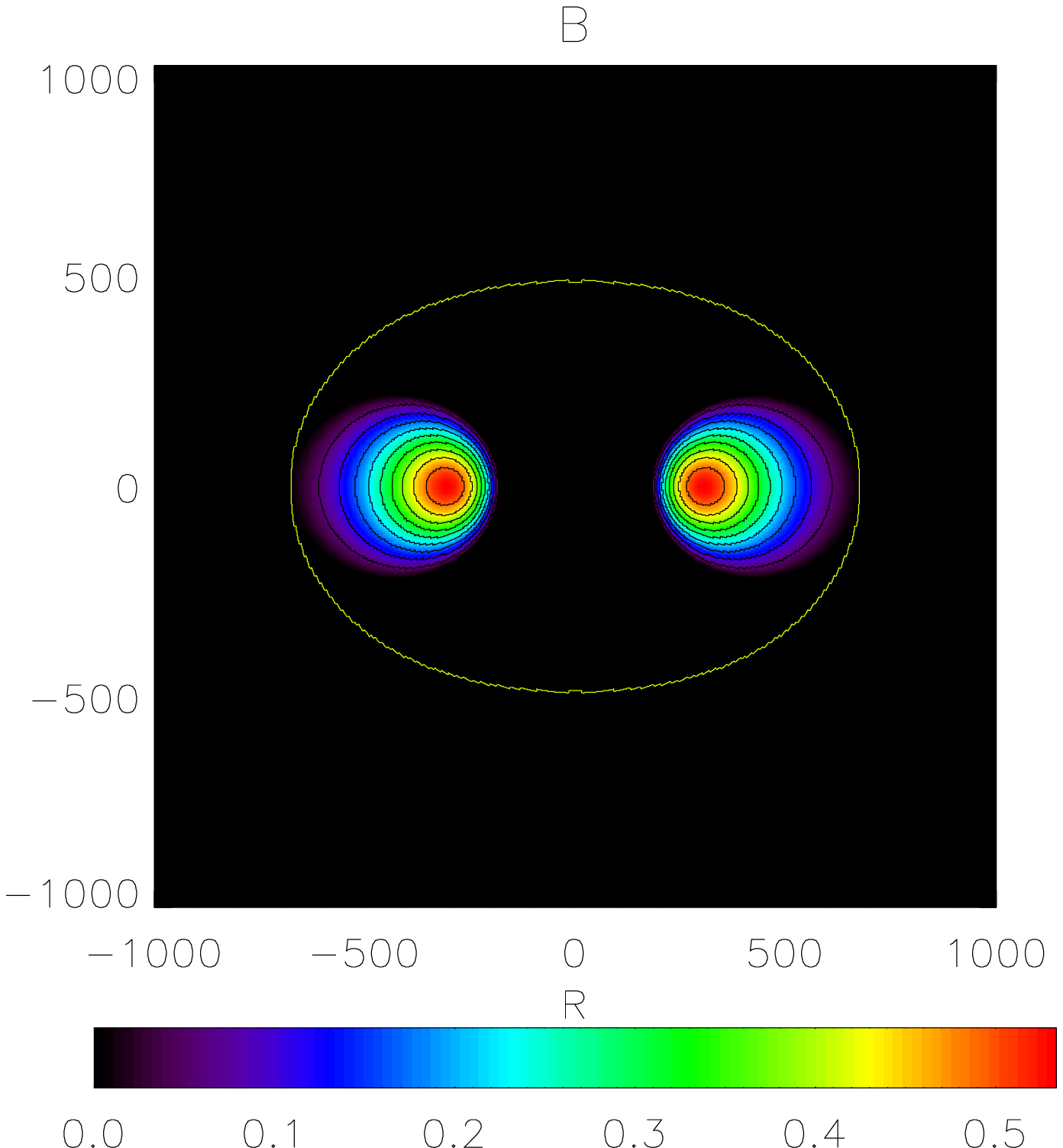}\figsubcap{a}}
  \hspace*{4pt}
  \parbox{2.1in}{\includegraphics[width=2in]{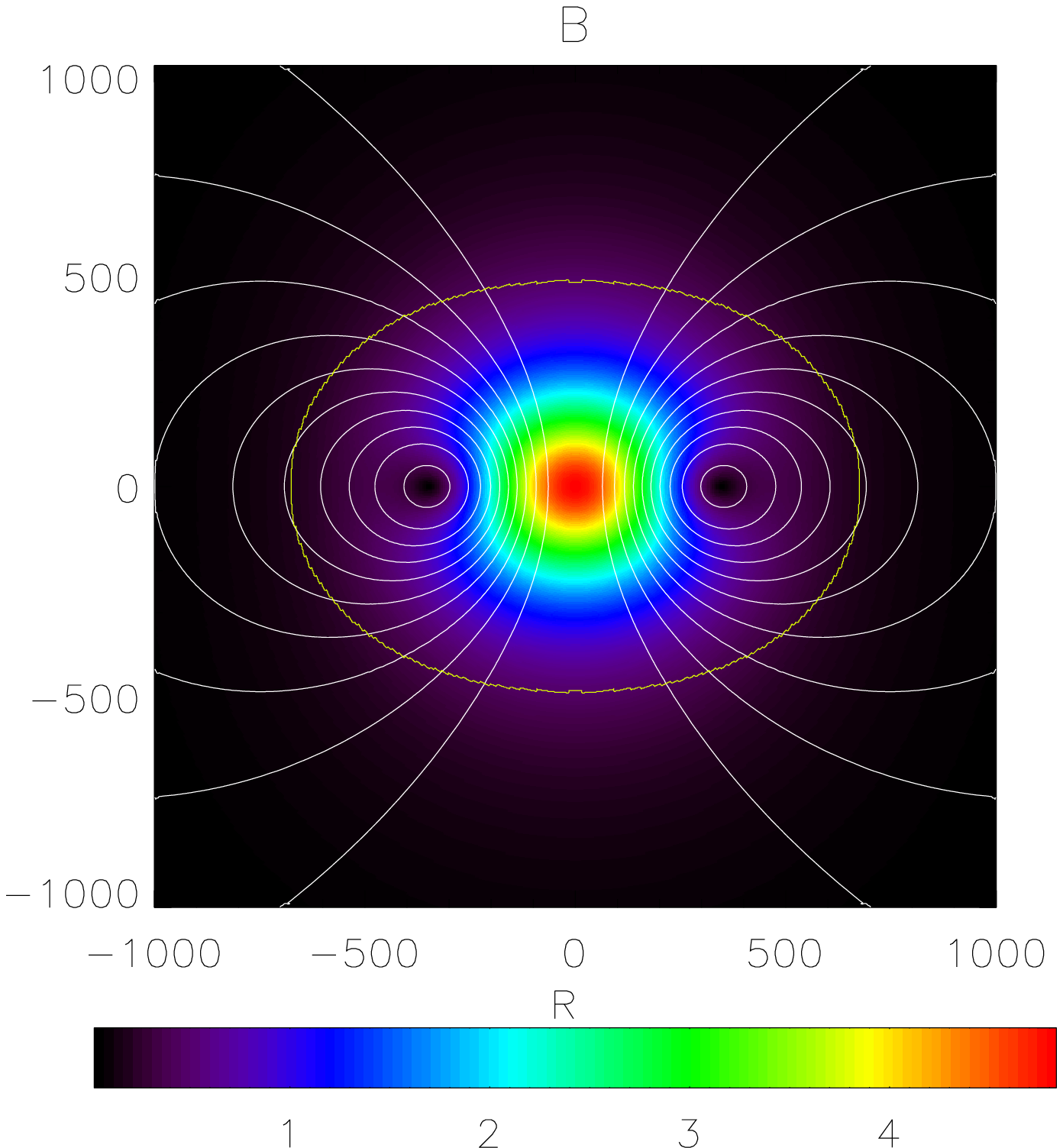}\figsubcap{b}}
  \caption{Mixed field or twisted torus configuration: magnetic field strength of (a) toroidal component and  (b) poloidal component.}%
  \label{mixed}
\end{center}
\end{figure}

Purely toroidal and poloidal field configurations 
are believed to be subjected to MHD instabilities,\cite{tayler} which 
possibly rearranges the field into a mixed configuration.\cite{braith} 
Hence, for completeness, we also construct equilibrium models of 
WDs with a mixed magnetic field configuration. 
In Figure \ref{mixed}, we present the results for the fiducial WD having the 
so-called {\it twisted torus configuration}, which compares the distribution 
of the toroidal and poloidal components of the magnetic field. 
This again results in a significantly super-Chandrasekhar WD 
having $M_0=1.754M_\odot$. 
The poloidal component attains $B_{\rm max}=4.82 \times 10^{14}$ G at the center, 
while the ring-like toroidal component is an order of 
magnitude smaller.
The WD assumes a highly 
oblate shape with $R_p/R_{eq} = 0.719$, resembling the purely poloidal 
case in all its attributes, and is again 
expected to be stable having $E_{mag}/E_{grav}=0.1126<1$. 
In a toroidal dominated 
mixed field configuration,\cite{ciolfi13} more massive super-Chandrasekhar WDs 
could be possible and is worth further exploration. 
In this context, we mention that the results in this work
have also been reproduced by Bera \& Bhattacharya,\cite{bera16} 
albeit without any novel contribution to the topic.


We also construct equilibrium sequences of magnetized WDs pertaining to different field 
geometries for the fiducial case with $\rho_c=2\times 10^{10}$ gm/$\rm cm^3$. 
Figure \ref{rhocf} shows the variations of different physical quantities as functions of $B_{\rm max}$.
The two most important revelations of Figure \ref{rhocf} are --- (1) the WD mass increases 
with an increase in magnetic field for all the 
three field configurations discussed above (see Figure \ref{rhocf}a), eventually leading 
to highly super-Chandrasekhar WDs and (2) the magnetic energy 
remains (significantly) sub-dominant compared to the gravitational binding energy for all the cases, 
since $E_{mag}/E_{grav}<1$ always (see Figure \ref{rhocf}d), which argues for the possible existence of stable, 
highly magnetized super-Chandrasekhar WDs.

\section{Conclusions}
\label{conc}

Since our foray into this topic, we have been persistent with our message 
that the versatile nature of magnetic field is 
paramount in the revelation of significantly super-Chandrasekhar WDs, irrespective of its 
nature of origin: quantum, classical and/or general relativistic --- which is re-emphasized 
in the present work.

By carrying out extensive, self-consistent, GRMHD numerical analysis of magnetized WDs, we have 
reestablished the existence of highly super-Chandrasekhar, stable WDs. 
In order to self-consistently study the anisotropic effect 
of a strong magnetic field, we have explored 
various geometrical field configurations, namely, purely toroidal, purely poloidal and twisted torus 
configurations. Interestingly, we have obtained significantly super-Chandrasekhar 
magnetized WDs for all the cases, having mass $1.7-3.4 M_\odot$, and that too
at relatively lower magnetic field strengths when the deviation 
from spherical symmetry is considered --- 
as already speculated in our earlier work.\cite{prd12,mg13} 
These WDs can be ideal progenitors of the aforementioned peculiar, overluminous type Ia supernovae.
Our work also establishes 
the necessity of a general relativistic formalism over a Newtonian 
approach while constructing models of magnetized super-Chandrasekhar WDs.

\vspace{-0.4cm}

\section*{Acknowledgments}

The authors thank N. Bucciantini and A.G. Pili for their inputs with the {\it XNS} code. 
This work was partly supported by the ISRO grant ISRO/RES/2/367/10-11 and was carried out 
at the Department of Physics, Indian Institute of Science, Bangalore, India.


\begin{thebibliography}{0}

\bibitem{prd12} U. Das and B. Mukhopadhyay, {\it Phys. Rev. D} {\bf 86}, 042001 (2012).


\bibitem{prl13} U. Das and B. Mukhopadhyay, {\it Phys. Rev. Lett.} {\bf 110}, 071102 (2013).


\bibitem{apjl13} U. Das, B. Mukhopadhyay and A.~R. Rao, {\it ApJ} {\bf 767}, L14 (2013).



\bibitem{jcap14} U. Das and B. Mukhopadhyay, {\it JCAP} {\bf 06}, 050 (2014).


\bibitem{kundu} A. Kundu and B. Mukhopadhyay, {\it MPLA} {\bf 27}, 1250084 (2012).

\bibitem{mg13} U. Das and B. Mukhopadhyay, {\it Proceedings of the Thirteenth Marcel Grossmann Meeting 
on General Relativity, July 2012}, 
(World Scientific Publishing, 2015).

\bibitem{chandra35} S. Chandrasekhar, {\it MNRAS} {\bf 95}, 207 (1935).




\bibitem{nature}
D.~A. Howell, et al. {\it Nature} {\bf 443}, 308 (2006).


\bibitem{scalzo}
R.~A. Scalzo, et al. {\it ApJ} {\bf 713}, 1073 (2010).

\bibitem{kepler} 

S.~O. Kepler, et al. {\it MNRAS} {\bf 429}, 2934 (2013).

\bibitem{grmhd}
U. Das \& B. Mukhopadhyay, {\it JCAP} {\bf 05}, 016 (2015).
















\bibitem{xns1} N. Bucciantini and L. Del Zanna, {\it A\&A} {\bf 528}, A101 (2011).

\bibitem{pili} A.~G. Pili, N, Bucciantini and L. Del Zanna, {\it MNRAS} {\bf 439}, 3541 (2014).


\bibitem{monica} M. Sinha, B. Mukhopadhyay and A. Sedrakian, {\it Nucl. Phys. A} {\bf 898}, 43 (2013).


\bibitem{ost} J.~P. Ostriker and F.~D.~A  Hartwick, {\it ApJ} {\bf 153}, 797 (1968).



\bibitem{tayler} R.~J. Tayler, {\it MNRAS} {\bf 161}, 365 (1973).

\bibitem{braith} J. Braithwaite, {\it MNRAS} {\bf 397}, 763 (2009).

\bibitem{ciolfi13} R. Ciolfi and L. Rezzolla, {\it MNRAS} {\bf 435}, L43 (2013).

\bibitem{bera16} P. Bera and D. Bhattacharya, {\it MNRAS} {\bf 456}, 3375 (2016). 



\end{thebibliography}
\end{document}